\documentclass{article}
\usepackage{ijcai26}

\usepackage{times}
\usepackage{soul}
\usepackage{url}
\usepackage[hidelinks]{hyperref}
\usepackage[utf8]{inputenc}
\usepackage[small]{caption}
\usepackage{graphicx}
\usepackage{amsmath}
\usepackage{amsthm}
\usepackage{booktabs}
\usepackage{algorithm}
\usepackage{algorithmic}
\usepackage[switch]{lineno}

\title{Compliance-Aware Agentic Payments on Stablecoin Rails}

\author{
Kenneth See$^1$\footnote{The views and opinions expressed in this paper are solely those of the authors and do not necessarily reflect the views of the Monetary Authority of Singapore and the Infocomm Media Development Authority.}
\and
Xue Wen Tan$^2$\footnotemark[1]\\
\affiliations
$^1$Monetary Authority of Singapore\\
$^2$Infocomm Media Development Authority\\
\emails
}
\begin{document}

\maketitle

\begin{abstract}
Agentic payment systems extend delegated action to financial transfers, but scaling them on stablecoin rails in regulated settings requires safeguards that remain effective when humans are not continuously in the loop. We present a compliance-aware architecture that combines x402-style, signature-based payment authorisation and relayed execution with programmable compliance embedded as an on-chain guardrail via a policy wrapper and policy manager coordinating modular checks. By enforcing compliance at the point of execution, rather than as a separate off-chain workflow, the approach preserves low-friction settlement when conditions are satisfied, records transaction-linked on-chain attestations, and supports structured resolution when requirements are pending.
\end{abstract}

\begin{figure*}[t]
    \centering
    \includegraphics[width=0.82\textwidth]{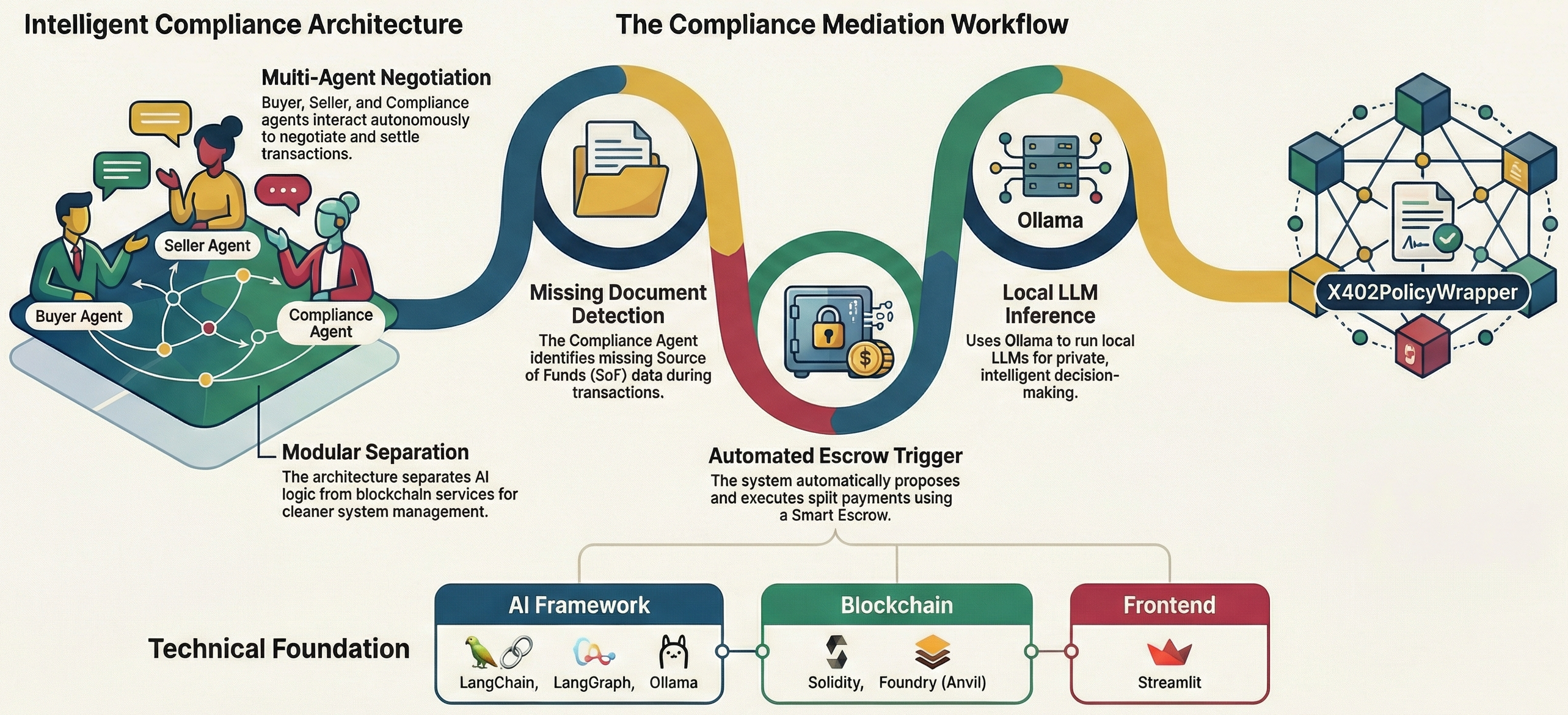}
    \caption{Solution overview}
    \label{fig:solution-overview}
\end{figure*}

\section{Introduction}
The past several years have witnessed the emergence of what is increasingly termed agentic artificial intelligence (AI), where systems are able to plan, reason, and execute multi-step actions rather than merely respond to isolated prompts. While such systems remain far from fully mature, recent industry evidence suggests a rapid rise in experimentation and early deployment. A global survey reports that more than one-third of organizations have already implemented agentic AI systems in some form, with a further large share actively planning adoption \cite{bcg_agentic_adoption_2025}. This accelerating interest is beginning to extend beyond productivity support toward delegated action in digital markets and financial operations. In emerging ``agentic commerce'' journeys, agents can search for products, negotiate terms, and execute purchases on behalf of users, while related exploratory work highlights their potential to support real-time treasury and cashflow decision processes in payment systems, together suggesting opportunities to improve conversion and reduce operational overhead for merchants and financial operators \cite{visa_agentic_commerce_2025,aldasoro_desai_bis_wp1310_2025}. Together, these developments motivate growing interest in agentic payments, where machines initiate and complete value transfers under user-specified constraints and at the speed and granularity expected of modern digital services.

However, scaling agentic payments requires confronting distinct risks introduced by autonomy, delegation, and multi-step tool use. First, agents can drift into technological paternalism, subtly substituting the agent's optimization for the user's autonomy, especially in decision support contexts where ``helpfulness'' is ill-defined \cite{mihale_wilson_invisible_paternalism_2025}. Second, agents are vulnerable to manipulation and hijacking in environments that contain untrusted content or tools. Surveyed threat classes include prompt and tool injection, environment tampering, and exfiltration pathways that can induce agents to reveal sensitive information or execute unintended actions \cite{deng_agents_under_threat_2025}. Third, even without adversaries, autonomy can lead to actions that deviate from a user's intent due to goal mis-specification, context loss, or brittle orchestration across tools and APIs \cite{leo_threat_to_trust_2026}. Consequently, responsible adoption requires safeguards that support governance, transparency, and regulation in ways that remain effective even when humans are not continuously in the loop \cite{murugesan_agentic_ai_2025}.

A second trend relevant to agentic payments is the growing adoption of stablecoins. Stablecoins are digital tokens designed to maintain a relatively stable value, typically through linkage to a fiat currency or reserve assets, and are commonly issued and transferred on blockchain networks that support smart contracts and programmable settlement \cite{higginson_spanz_stable_door_2025}. For certain payment flows, stablecoins can offer faster settlement, broader availability across borders and time zones, and potentially lower operational friction than legacy rails, especially where intermediated reconciliation and batch settlement dominate \cite{higginson_spanz_stable_door_2025}. These properties make stablecoins a plausible substrate for agentic payments, where agents may need to transact frequently, automatically, and in near real time \cite{x402_whitepaper_2025}.

Yet, payments remain a highly regulated activity, and compliance does not disappear simply because payment rails change. Conventional requirements include sanctions screening, customer due diligence controls, recordkeeping, and information-sharing obligations such as the FATF ``travel rule'' (Recommendation 16) as applied to virtual assets and VASPs \cite{fatf_targeted_update_2023}. Regulators also emphasize that faster or instant payment schemes can intensify sanctions-compliance challenges by compressing the time available to interdict prohibited transfers, and therefore should incorporate sanctions controls by design \cite{ofac_instant_payments_guidance_2022}. More broadly, introducing distributed ledger infrastructure can reconfigure how trust is preserved and evidenced across participants, but it does not eliminate the need to demonstrate compliance and accountability \cite{see_li_permissioned_blockchains_trust_2025}. If compliance remains primarily manual, slow, or externally mediated, the purported benefits of real-time agentic payments and stablecoin-based settlement are unlikely to materialize at scale.

This motivates interest in programmable compliance, which leverages smart-contract affordances to encode compliance requirements as enforceable conditions for transfers. Prior work has highlighted how ``conditional payments'' and policies attached to money or transfers can be represented and enforced on-chain \cite{weber_staples_programmable_money_2022}. In practice, multiple industry efforts are converging on architectures for regulated digital assets that embed identity, permissions, and policy checks into token transfer logic or associated rails, including permissioned token standards (e.g., ERC-3643) \cite{tokeny_erc3643_whitepaper_2023}, payment-token design proposals for safety and integrity \cite{toh_mit_dci_kinexys_2025}, and emerging reference architectures for programmable compliance that distinguish non-interactive checks from interactive, evidence-producing workflows \cite{gl1_pc_toolkit_2025}. Complementarily, x402 proposes an HTTP-native payment flow that allows software clients, including agents, to programmatically pay for services using stablecoins, commonly leveraging EIP-3009 \texttt{transferWithAuthorization} to support relayed execution via signatures \cite{eip3009_2020,x402_whitepaper_2025}.

In this paper, we demonstrate how programmable compliance integrated into stablecoin payment rails can support agentic payment systems that are both seamless and regulated. Concretely, we demonstrate an agentic workflow in which a compliance mediator coordinates buyer and seller agents, executes on-chain policy evaluation, and supports interactive resolution when evidence (e.g., source-of-funds attestation) is missing. The design preserves a machine-native execution path for transfers while preventing unauthorized or non-compliant settlement, thereby reducing compliance-related friction that would otherwise block real-time agentic payments. Our results provide a practical pathway for scaling agentic payments by making regulatory safeguards enforceable, inspectable, and automatable, without requiring the user experience to revert to manual, out-of-band compliance checks.
\section{Architecture and Implementation}

Our system is a modular, hybrid architecture that couples agentic decision-making with deterministic on-chain enforcement (Figure~\ref{fig:solution-overview}). We assume a simple commerce transaction in which a buyer agent expresses purchase intent, a seller agent provides inventory and pricing, and a compliance agent mediates payment execution under a programmable compliance regime\footnote[2]{Implementation code can be found at \url{https://github.com/Chen-XueWen/agentic_compliance_payment}}.%\footnote{Our Github repository will be made available upon publication.}.

At a high level, the system comprises (i) a lightweight user interface for issuing intents and observing execution, (ii) an agent orchestration layer that coordinates the buyer, seller, and compliance agents, and (iii) an on-chain enforcement layer realized through the Global Layer One (GL1) Programmable Compliance architecture. The agents interact with the on-chain layer via an x402-style gateway: the buyer produces a signed payment authorization, and the compliance agent compiles and submits the corresponding payment instruction to the GL1 policy wrapper for evaluation and settlement.

Compliance outcomes are returned as transaction-linked attestations (\texttt{PASS}/\texttt{FAIL}/\texttt{PENDING}). When checks pass, settlement proceeds immediately; when requirements are pending, the compliance agent can negotiate a structured alternative (e.g., staged settlement) and complete payment once the required evidence is provided. This design embeds compliance as a guardrail at the point of execution, avoiding the need to treat agentic workflows and payment compliance as separate, friction-inducing processes.

%[TO-DO]
%\begin{itemize}
%    \item Leverage x402 payment protocol for agents to per payment discovery, payment authorization, and relay instructions for on-chain settlement
%    \item We implement a lite version of the Global Layer One Programmable Compliance reference model to support transactional compliance in our demo
%    \item We deploy the programmable compliance contracts and stablecoin on a local Ethereum Virtual Machine (EVM)-compatible development network for experimental evaluation
%    \item To discuss the roles and capabilities given to the agents (highlight negotiation)
%    \item Each agent is associated with an on-chain wallet address, which serves as the on-chain identifier and settlement account for the party it represents
%\end{itemize}

\section{Demo}

In the absence of programmable compliance safeguards, naïve agentic payment flows can lead to ethically and institutionally problematic outcomes. Transactions may settle despite violating regulatory constraints, exposing counterparties to compliance risk, while informal, off-chain compliance handling can incentivize excessive disclosure of identity data, undermining privacy and data-minimization principles \cite{salmon_agentic_ai_finance_2025}. These risks illustrate how unstructured agentic payments can erode both regulatory trust and user autonomy.

The following subsections present two demo scenarios that illustrate how embedding programmable compliance into the payment substrate reshapes these flows. The first scenario shows a compliance-aware payment that settles immediately when regulatory conditions are satisfied, while the second demonstrates how pending compliance requirements are surfaced and resolved through agent-mediated tranching and escrow.

\subsection{Scenario I: Compliance-Aware Agentic Payment Flow}

This scenario illustrates a compliance-aware agentic payment in which regulatory requirements are already satisfied, allowing settlement to proceed seamlessly while producing on-chain compliance attestations. The buyer agent initiates the transaction by generating a signed payment authorization, which is submitted on-chain by a compliance agent acting as a mediator. Rather than executing a direct transfer, the authorization is routed through the \texttt{PolicyWrapper}, which serves as the x402-compatible entry point to the programmable compliance framework.

\begin{figure}[hbt!]
    \centering
    \includegraphics[width=\columnwidth]{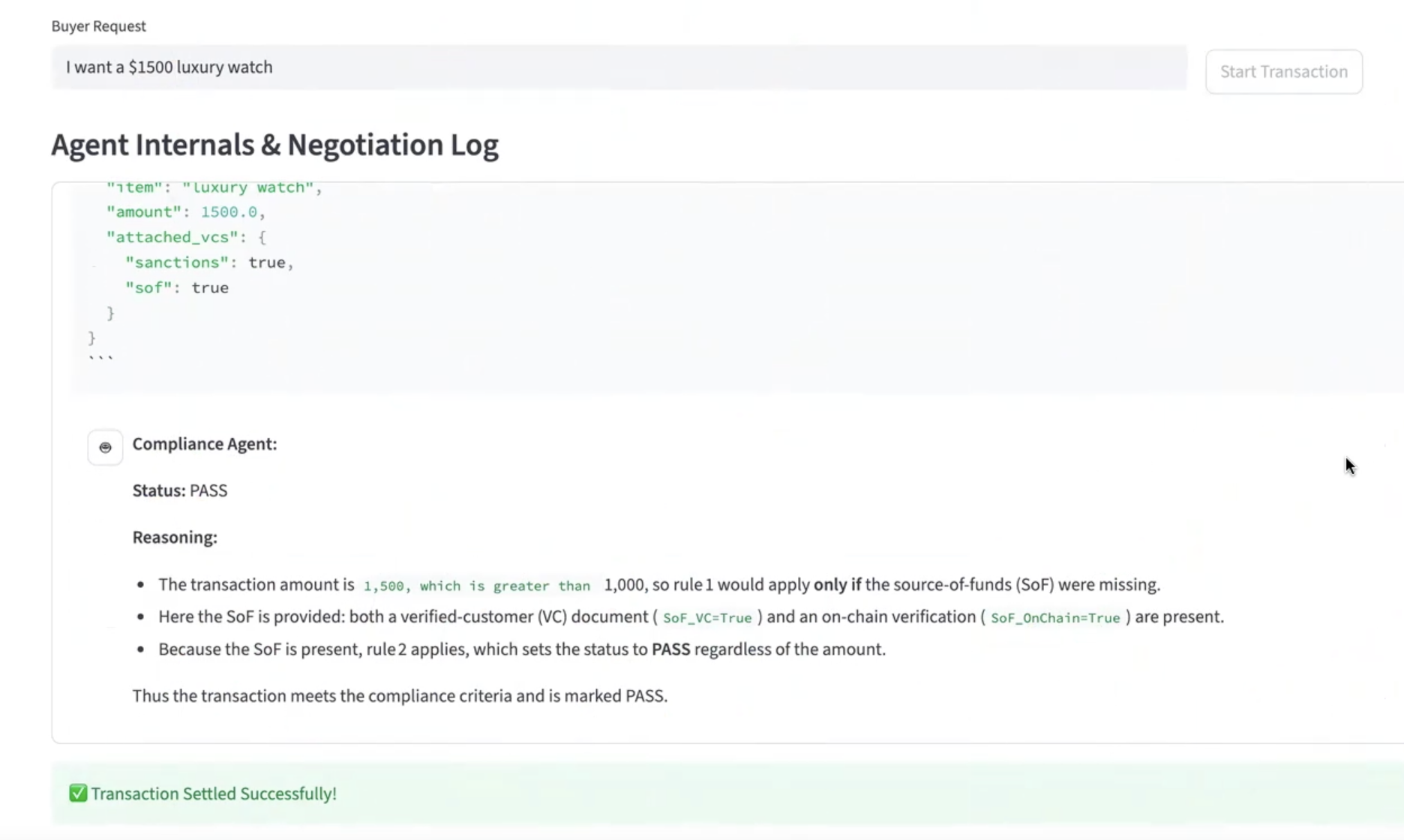}
    \caption{User interface illustrating a compliance-aware agentic payment flow, including the attestation checks and eventual transaction completion}
    \label{fig:scenario1-ui}
\end{figure}

The \texttt{PolicyWrapper} evaluates the transaction against the relevant sanctions and source-of-funds policies. In this case, all policy checks return a \texttt{PASS} result, triggering immediate settlement without additional interaction or delay (Figure~\ref{fig:scenario1-ui}). A compliance attestation linked to a unique transaction identifier is recorded on-chain, together with updated account balances, demonstrating how compliance enforcement can be embedded into agentic payment flows without introducing additional friction.

\subsection{Scenario II: Pending Compliance and Escrow-Mediated Settlement}

This scenario demonstrates how the system handles transactions that cannot be immediately approved. When a payment exceeds the source-of-funds threshold, on-chain policy evaluation returns a \texttt{PENDING} result. No funds are settled; instead, a compliance attestation is recorded on-chain indicating that additional evidence is required.

\begin{figure}[hbt!]
    \centering
    \includegraphics[width=\columnwidth]{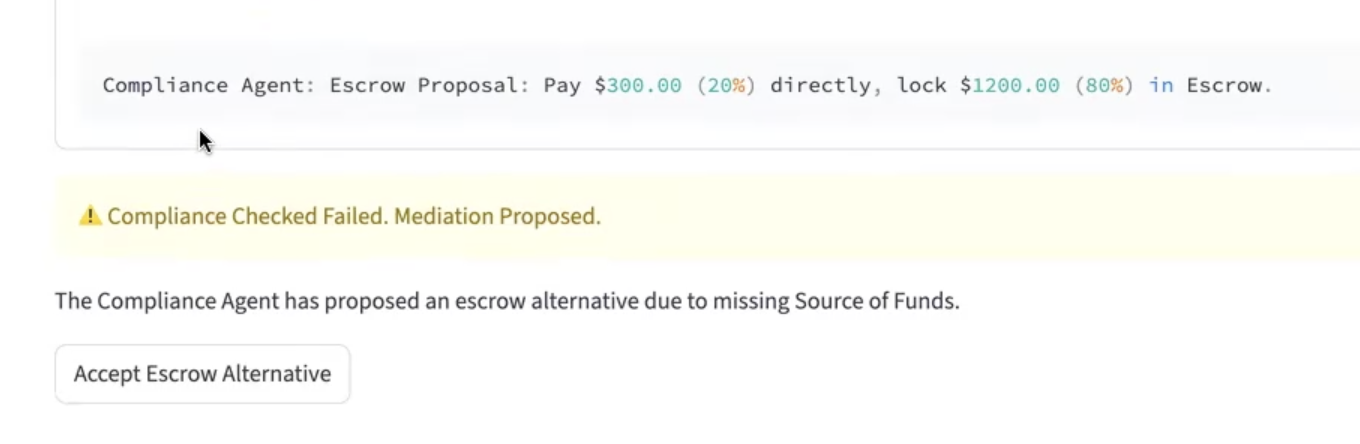}
    \caption{User interface illustrating compliance failure and alternative proposal by the compliance agent}
    \label{fig:scenario2-ui}
\end{figure}

The compliance agent retrieves the pending result and proposes an alternative arrangement in which the payment is split into two tranches (Figure~\ref{fig:scenario2-ui}). An initial tranche that satisfies existing conditions is settled, while the remaining amount is locked in an escrow contract pending resolution. The buyer subsequently submits a source-of-funds attestation, which is recorded on-chain and linked to the buyer’s identity. Once the requirement is satisfied, the compliance agent releases the escrowed funds, completing settlement. This scenario highlights how incomplete compliance conditions can be resolved through agent-mediated workflows without reverting to manual or off-chain intervention.

\section{Conclusion}

We present an agentic payment system that integrates programmable compliance directly into stablecoin payment rails, demonstrating how regulatory safeguards can be enforced without undermining the seamless execution expected of agentic systems. Through compliance-aware and escrow-mediated scenarios, we showed how on-chain policy evaluation enables immediate settlement when requirements are satisfied, while supporting structured resolution when regulatory conditions remain pending. By embedding compliance logic into the payment substrate and aligning it with x402-style authorization flows, our approach illustrates a practical pathway to realizing the benefits of agentic payments—automation, speed, and machine-native interaction—while preserving regulatory trust and minimizing friction.
%%%%%%%%%%%%%%%%%%%%%%%%%%%%%%%%%%%%%%%%%%%%%%%%%%%%%%%%%%%%%%%%%%%%%%%%%%%
%%%%%%%%%%%%%%%%%%%%%%%%%%%%%%%%%%%%%%%%%%%%%%%%%%%%%%%%%%%%%%%%%%%%%%%%%%%

\newpage
\section*{Contribution Statement}
All authors contributed equally to the conception, design, analysis, and writing of this paper.

\bibliographystyle{named}
\bibliography{ijcai26}

\end{document}